# Statistical Thermodynamic Foundation for the Origin and Evolution of Life

Ronald F. Fox
Regents' Professor Emeritus
School of Physics
Georgia Institute of Technology
Atlanta, Georgia

## Abstract

In this paper we review and extend our earlier recent work on thermostated systems. A description of nano-biological systems by Markov chains in coordinate space in the strongly overdamped limit is presented. Characterization of the most probable path is given and a new formula for the probability of this special path is provided from recursion formulae. The deterministic limit is derived and the significance of Lagrange multipliers introduced when constructing the most probable path is elucidated. The characterization of the generation of path entropy by the most probable path is given an equivalent interpretation relating to the rate of entropy production by the most probable path. The paper concludes with an account of the biological implications. Here we address *why* the origin of life and its subsequent evolution took place, not the particular chemical details of *how* it happened.

## I. Introduction

In his insightful book, *The Logic of Life* [1], F. Jacob addresses the question: what is the minimal structure that can be called living? He defines such a structure to be one that minimally supports its own existence. He argues that the answer is a simple cell, a membranous structure containing all the necessary enzymes for energy metabolism and polymer synthesis, including genetic material that can be transcribed, translated and replicated. In the very early stages of the emergence of life such a structure, a *protocell*, would have to harness and process energy to be able to manufacture the polymers needed for function. A recent review of this viewpoint is in a paper by D. Deamer and A. L. Weber [2]. This conclusion is in sharp contrast with the RNA-world view in which the minimal structure is a self-replicating ribonucleic acid polymer. A recent proponent of this view is H. S. Bernhardt [3]. Earlier than either of these two reviews, C. De Duve wrote a comprehensive analysis of the pros and cons of both views in a book about the origin of life [4]. He came down strongly on the side of the protocell and emphasized that energy harnessing and processing were the key events around which all other developments in evolving life were organized. He recognized a sequence of energy transductions starting from light driven redox reactions in iron atoms, through sulfur-dependent thioesters, to polyphosphates such as ATP (adenosine triphosphate). This view recalls the earlier thinking of F. Lipmann [5] and of A. Szent-Gyorgyi [6]. Regardless of which side one takes, there remains a basic question: *what is the physical reason that natural processes move in the direction of the origin and evolution of life?* Why don't they instead take molecules to their lowest free energy states quickly and completely? Even though the flow of redox energy, its coupling to the production of thioesters and then, by means of the thio-phosphate intermediate, its formation of polyphosphates, will drive monomeric molecules out of their lowest free energy states by phosphorylation (activation), why do these activated monomeric molecular states lead to the emergence of either protocells or self-replicating RNA molecules instead of simple cycles of activation and decay (hydrolysis)? In this paper we will argue that statistical thermodynamics in the very special context appropriate for the emergence of life at the nano-scale (nanometers and nanoseconds) provides a putative answer. Quite simply put, these emergent complexities are the *most probable* events.

Energy metabolism and the monomer to polymer transition

Cellular life is characterized by the central importance of three major classes of polymers, the proteins, the polynucleotides and the polysaccharides (there are also the fatty acid hydrocarbon chains the synthesis of which involves the dehydration mechanisms referred to below but does so in a much more chemically complex way that involves many thioesters). These macromolecules are synthesized inside cells from three classes of small molecules called monomers and they are respectively the *amino acids*, the *mononucleosides* and the *monosaccharides*. Moreover the transition from monomers to polymers, the chemical hallmark of life, follows one universal mechanism, *dehydration condensation.* This means that the bond joining monomers, respectively the *peptide bond*, the *phosphodiester bond* and the *glycosidic bond*, results in the release of a molecule of water, at least that is the situation for a spontaneous dehydration. The problem with this mechanism is that the cell is largely made of water so that a mechanism releasing water is thermodynamically inhibited by end product inhibition. In water, spontaneous dehydrations are extremely rare. Hydrolysis rules instead. Consequently the cell must provide *activation energy* to circumvent this inhibition. This it does by means of a universal energy carrier and activator the polyphosphate. Today almost all organisms use ATP for this purpose, forming an activated phosphate intermediate, respectively the *amino acyl adenylate*, the *nucleotide triphosphate* and the *uridine diphospho monosaccharide*. The production of polymers requires large-scale production of ATP in order to generate these phosphorylated activated monomers. The core of metabolism is designed around this requirement and is called energy metabolism. It includes glycolysis (inefficiently generates ATP from oxidation of glucose but does it rapidly), the acetyl-CoA bridge (connects glycolysis to the citric acid cycle and to fatty acid generation), the citric acid cycle (generates electron reducing potential, NADH, and GTA, an ATP cousin) and in aerobes, the electron transport chain (converts electron reducing potential into ATP slowly but very efficiently, and utilizes proton current (chemiosmosis) to couple reducing electron current to ATP synthesis). A lot of enzymes are involved in membrane bound complexes. They are translated from a lot of genes. These thoughts make one very dubious about a lone RNA molecule having had the evolutionary capacity to fulfill all these requirements by itself. Therefore, RNA-world advocates are faced with the daunting task of imagining how a self-replicating RNA molecule managed to evolve the needed structures to eventually become a primitive cell with energy metabolism and proteins, and the enzymes needed to generate its monomeric precursors. Very clever researchers have made progress with parts of this challenge but we may still ask why would RNA make these advances rather than to simply hydrolyze into monomeric matter in harmony with normal thermodynamics? It is the general version of this question that we will address below.

Why do natural processes go in the direction of life?

Spontaneous natural processes in closed systems progress in the direction of monotonic decrease in the Gibbs free energy at constant temperature and pressure. Here we contemplate processes that are driven from outside either by UV excitation of iron electrons or by other mechanisms that can produce reducing equivalents of electrons (e.g. heat). These activated electrons can drive the formation of polyphosphates (via thioesters), as was alluded to above, and energy in this form, polyphosphates, is ideally suited for the activation of each class of monomers and their subsequent polymerization (including triglyceride synthesis). In order to focus on the central theme of this paper we will imagine starting with a mixture of polyphosphates and monomers and ask the question what happens to this closed system as the monotonic decrease in Gibbs free energy is followed in time. One possibility, the most intuitively likely, it that the polyphosphates rapidly hydrolyze and we quickly reach the equilibrium thermodynamic state of unpolymerized monomers and unpolymerized phosphates, coupled to a release of heat. Another possibility is that the polyphosphates activate the monomers, with a small decrease in free energy, followed by the activated monomers polymerizing into polymers, again with small decreases in free energy, and finally the slow hydrolysis of the generated polymers. The end result is the same, an equilibrium mixture of monomers and phosphate, but the approach has been slow and the intermediate polymers have had a chance to possibly affect various types of feedback effects on the evolving chemical mixture. Why would the second possibility dominate the first? A chemist would answer that it has to do with rates and the monotone decrease in free energy is a manifestation of the Second Law that says nothing about absolute rates, only having implications for equilibrium and rate ratios. When we distinguish between global thermodynamics for a system (total temperature, volume and mole numbers) and the local statistical thermodynamics for the time dependent stochastic paths in the system, we find that rates do come into the picture after all.

Outline of paper

The paper is organized as follows. In section **II** we review and extend our earlier recent work on thermostated systems [7a, b]. We show how to describe nano-biological systems by Markov chains in coordinate space in the strongly overdamped limit. Characterization of the most probable path is given and a new formula for the probability of this special path is provided from recursion formulae. In section **III** we consider the deterministic limit and the significance of Lagrange multipliers introduced when constructing the most probable path. In section **IV** we characterize the generation of path entropy by the most probable path and determine an equivalent interpretation relating to the rate of entropy production by the most probable path. In section **V** the biological implications are elucidated. Here we address *why* the origin of life and its subsequent evolution took place, not the particular chemical details of *how* it happened. This section is tentative given the complexity of the description.

## II. Review of thermostated systems

Thermostated systems are isothermal systems in which the temperature is a very slowly changing variable on the time scale for molecular events (sub-picoseconds) or even strictly constant. Otherwise such systems may be far from equilibrium in other quantities such as numbers of molecules of different types and the motion of these molecules. In the context of nano-biology emphasized in this paper the source of temperature stability is the interstitial water that is a very good heat conductor with a large heat capacity, features that make for a very good thermostat. Over the past twenty years much research has focused on the behavior of thermostated systems in phase space (coordinates and momenta) [8-13]. Because computers can be used in these studies it is possible to instantaneously stop a trajectory and reverse each and every momentum variable and then evolve the system forward in time starting from the reversed momenta. Comparison of the forward and reverse probabilities for these trajectories has lead to several insights [10, 11]. D. Andrieux and P. Gaspard [14] and also J. L. English [15] have broached the possibility that these insights have significance in nano-biology, as have I [7b].

In my approach to thermostated systems, I eliminate the explicit momenta dependence by contracting the description into a solely coordinate dependent picture. I do this because the molecular momenta are randomly changing direction on a very short time scale (less than picoseconds) such that the effect of all the water molecules is to produce Brownian motion on all of the solute species. Thus, the explicit presence of water is replaced by stochastic forces that create Brownian motion in a very low Reynolds number environment. On the nano-scale the dynamics can be said to be strongly overdamped and inertia plays no role whereas viscosity is overwhelmingly dominant. The elimination of momenta is achieved using projection operator techniques and yields a non-Markov process. In the strongly overdamped regime, this process is very well approximated by a Markov process. The details of these steps in the derivation of the Markov contracted description are in [7a, b].

A consequence of the procedure outlined above, with its Markov outcome, is that the analysis in full phase space can be paralleled in the reduced, or contracted, coordinate space. In our recent work with this description [7b], we showed that the most probable path between two points in coordinate space, for a fixed time of transit, minimizes the entropy generated, or minimizes the decrease in Helmholtz free energy. While at first sight this appears to be in conflict with J. England's findings in full phase space [15], D. Ruelle has argued that it is not [16] because of the basic difference in context.

## Markov chains and the most probable path

In [7b] the equation for the Markov contracted description in one coordinate variable and a potential $U(r)$ is given by

(1)
$$\partial_t R(r,t) = D\partial_r(\partial_r + \beta U')R(r,t)$$

in which $R(r,t)$ is the probability density for coordinate $r$ and $\beta = \frac{1}{k_B T}$ where $T$ is the temperature and $k_B$ is Boltzmann's constant. It is easy to see that the Boltzmann distribution is the equilibrium solution to this equation, and in [7b] it is shown that this macroscopic equilibrium is approached monotonically. In this one coordinate case $U(r)$ is the container potential. Generalization the $n$ particles, each in three dimensions, is straight forward and yields the equation

(2)
$$\partial_t R = \sum_{l=1}^{n} D_l \nabla_l \cdot (\nabla_l + \beta \nabla_l U) R$$

in which $R = R(r_1, r_2, \ldots, r_n, t)$ and the possibility that molecules of different types have different diffusion constants is manifested by their subscripts. In this expression $U(r)$ includes the container potentials as well as the interparticle potentials that may involve pairs and higher order interactions.

(3)
$$U = \sum_{j=1}^{n} U_j + \frac{1}{2}\sum_{j,k=1}^{n} U_{jk} + \frac{1}{6}\sum_{j,k,m=1}^{n} U_{jkm} + \cdots$$

It is shown in [7b] that the Green's function solution to Eq.(1) is also the conditional probability distribution for the Markov process,

$P_2(\boldsymbol{r}^{3n}(j), \boldsymbol{r}^{3n}(j-1), t)$ wherein a shorthand notation for the 3-d components of $n$ particles at time 0 is given by $\boldsymbol{r}^{3n}(j-1)$ and at time $t$ by $\boldsymbol{r}^{3n}(j)$. Ordinarily $j$ and $j-1$ would be subscripts, or possibly superscripts, but other labels will occupy these script positions instead. The initial value condition for the Green's function is

(4)
$$P_2(\boldsymbol{r}^{3n}(j), \boldsymbol{r}^{3n}(j-1), 0) = \prod_{l=1}^{n} \delta\left(\boldsymbol{r}_l(j) - \boldsymbol{r}_l(j-1)\right)$$

In general no closed form solution for the Green's function is obtainable, but for sufficiently short time intervals, $\Delta t$, the Trotter product formula can be used [7b] to obtain

(5)
$$P_2(\boldsymbol{r}^{3n}(j), \boldsymbol{r}^{3n}(j-1), \Delta t) =$$
$$\prod_{l=1}^{n} \left(\frac{1}{\sqrt{4\pi D_l \Delta t}}\right)^3 \exp\left[-\frac{(\boldsymbol{r}_l(j) - \boldsymbol{r}_l(j-1) + \beta D_l \Delta t \nabla_l U(\boldsymbol{r}_l(j-1)))^2}{4 D_l \Delta t}\right]$$

wherein the derivative of the potential is with respect to components of $\boldsymbol{r}_l(j-1)$. The correct way to interpret this notation is that at time 0 the value of the coordinate for particle with label $l$ is $\boldsymbol{r}_l(j-1)$ and at time $\Delta t$ its value is $\boldsymbol{r}_l(j)$.

We may contemplate a coarse-grained tube of paths in coordinate space with forward probability given by

(6)
$$P^F(path) = \prod_{j=1}^{N} d^{3n}\boldsymbol{r}(j) P_2(\boldsymbol{r}^{3n}(j), \boldsymbol{r}^{3n}(j-1), \Delta t)$$

On the right-hand side the conditional probabilities are given by Eq.(5) and the differential intervals are of fixed size for each particle type and for each value of $j$. The differentials convert probability densities into probabilities. The initial point in coordinate space for this path is given by $\boldsymbol{r}_l(0)$ for each $l$, and the final point of the path is given by $\boldsymbol{r}_l(N)$ for each $l$.

The most probable path connecting the initial and final points by $N$ very small steps is given by making two observations. Firstly, note that

(7)
$$\max \ln(P^F(path)) =$$
$$\max \sum_{j=1}^{N} \sum_{l=1}^{n} \left( \ln\left(\frac{d^3 r_l(j)}{(4\pi D_l \Delta t)^{3/2}}\right) - \frac{(r_l(j) - r_l(j-1) + \beta D_l \Delta t \nabla_l U(r^{3n}(j-1)))^2}{4 D_l \Delta t} \right)$$

Because all of the coordinately differentials are held fixed for the entire path the logarithmic term on the right-hand side may be dropped. The remaining term is negative so that, secondly, note that

(8)
$$\sum_{j=1}^{N} \sum_{l=1}^{n} \left( \frac{(r_l(j) - r_l(j-1) + \beta D_l \Delta t \nabla_l U(r^{3n}(j-1)))^2}{4 D_l \Delta t} \right) \text{ is a minimum}$$

subject to the constraint

(9)
$$r_l(N) - r_l(0) = constant \; for \; each \; l$$

The minimization, subject to the constraint, can be performed by introducing $n$ 3-d Langrange multiplier vectors, $\Lambda_l$. For each particle labeled by $l$, we minimize the quantity in Eq.(8) plus the Lagrange constraint expression,

(10)
$$\sum_{l=1}^{n} \Lambda_l \cdot (r_l(N) - r_l(0))$$

by differentiating the sum of Eq.(8) and expression (10) with respect to $r_l(j)$ for each $j$ from $j - 1 = 0$ to $j = N$. For each $l$ and for $j = N$ we get

(11)
$$\frac{1}{2 D_l \Delta t} (r_l(N) - r_l(N-1) + \beta D_l \Delta t \nabla_l U(r^{3n}(N-1))) + \Lambda_l = 0$$

Note that both terms on the left-hand side must have the units of inverse length. This expression is relatively simple because $r_l(N)$ only appears in the $j = N$ term. The next expression for $r_l(N-1)$ is more complicated because $r_l(N-1)$ appears in both the $j = N$ and $j = N-1$ terms and, for $j = N$, inside the potential as well. On the other hand there is no $r_l(N-1)$ in expression (10). For $j = N-1$ we get

(12)

$$\frac{1}{2D_l\Delta t}(r_l(N-1) - r_l(N-2) + \beta D_l \Delta t \nabla_l U(r^{3n}(N-2))) +$$

$$\frac{1}{2D_l\Delta t}(r_l(N) - r_l(N-1) + \beta D_l \Delta t \nabla_l U(r^{3n}(N-1))) \cdot \nabla_l$$

$$(r_l(N) - r_l(N-1) + \beta D_l \Delta t \nabla_l U(r^{3n}(N-1))) = 0$$

The gradients, $\nabla_l$, are with respect to $r_l(N-1)$. Since $r_l(N-1)$ and $r_l(N)$ are independent variables, the contribution of $r_l(N)$ to the last line of Eq.(12) is 0 and the contribution of $-r_l(N-1)$ is $-1$. This leaves just the potential term of the last line multiplied by either $r_l(N) - r_l(N-1)$ or by another potential term. The product of the potential terms is proportional to $\Delta t^2$ and will be dropped for very small $\Delta t^2$. The probability density in Eq.(5) shows us that the factor $r_l(N) - r_l(N-1)$ is of order $\Delta t^{1/2}$ or else has very small probability, so that its product with the potential term is order $\Delta t^{3/2}$ which we will also neglect. (To derive Eq.(12) the vector calculus identity $\nabla A \cdot A = 2A \cdot \nabla A + 2A \times \nabla \times A$ was used. The curl term makes no contribution.) In [7b] we showed that for reasonable values of the quantities in these expressions, $r_l(j) - r_l(j-1)$ is of order $\Delta t^{1/2}$ with a value of $4.5 \times 10^{-10}$ cm, and the complete potential term is manifestly of order $\Delta t$ with a value of $2.5 \times 10^{-12}$ cm. Putting this all together, Eq.(12) becomes

(13)

$$\frac{1}{2D_l\Delta t}(r_l(N-1) - r_l(N-2) + \beta D_l \Delta t \nabla_l U(r^{3n}(N-2))) -$$

$$\frac{1}{2D_l\Delta t}(r_l(N) - r_l(N-1) + \beta D_l \Delta t \nabla_l U(r^{3n}(N-1))) = 0$$

Eq.(11) implies

(14)

$$\frac{1}{2D_l \Delta t}(r_l(N-1) - r_l(N-2) + \beta D_l \Delta t \nabla_l U(r^{3n}(N-2))) = -\Lambda_l$$

Proceeding in this way from $j = N-1$ to $j = 2$ we get

(15)
$$\frac{1}{2D_l \Delta t}(r_l(j) - r_l(j-1) + \beta D_l \Delta t \nabla_l U(r^{3n}(N-1))) = -\Lambda_l$$

For just $j = 1$ does the $r_l(0)$ term appear, but it also does appear in the potential term and in the Lagrange multiplier expression (with a minus sign). The result is

(16)
$$\frac{1}{2D_l \Delta t}(r_l(1) - r_l(0) + \beta D_l \Delta t \nabla_l U(r^{3n}(0))) + \Lambda_l = 0$$

These recursion formulas can be used in two ways that were not anticipated in [7b].

A formula for the probability of the forward path can be obtained from the recursion formulas and a deterministic limit can be taken. Place Eq.(15) inside eq.(5) and then place the result inside eq.(6) to get

(17)
$$P^F(path) = \prod_{j=1}^{N} d^{3n}r(j) \prod_{l=1}^{n} \left(\frac{1}{\sqrt{4\pi D_l \Delta t}}\right)^3 \exp[-D_l \Delta t \Lambda_l \cdot \Lambda_l] =$$

$$\prod_{j=1}^{N}\prod_{l=1}^{n} d^3 r_l(j) \left(\frac{1}{\sqrt{4\pi D_l \Delta t}}\right)^3 \exp[-D_l \Delta t \Lambda_l \cdot \Lambda_l]$$

Replacing the left-hand side of Eq.(15) by the Lagrange vector on the right-hand side eliminates the $j$-dependence. Because we chose a tube of paths for which the differentials are the same for each $j$-step, we can replace $d^3 r_l(j)$ by $d^3 r_l(*)$ for each $j$. The product over $j$ can be performed and yields

(18)
$$P^F(path) = \prod_{l=1}^{n}(d^3 r_l(*))^N \left(\frac{1}{\sqrt{4\pi D_l \Delta t}}\right)^{3N} \exp[-D_l N \Delta t \Lambda_l \cdot \Lambda_l]$$

The factor $N\Delta t$ is the total time to go from the initial point in coordinate space to the final point in coordinate space. We may call this time interval $t_{if}$. This gives the final expression for the forward path probability

(19)
$$P^F(path) = \prod_{l=1}^{n}(d^3\boldsymbol{r}_l(*))^N\left(\frac{1}{\sqrt{4\pi D_l \Delta t}}\right)^{3N} exp[-D_l t_{if} \boldsymbol{\Lambda}_l \cdot \boldsymbol{\Lambda}_l]$$

### III. Deterministic limit and Lagrange multipliers

We are now ready to determine the Lagrange multipliers and the transit time. Several properties of the path expressions need to be clarified. The initial point, $\{\boldsymbol{r}_l(0)\}$, and the final point, $\{\boldsymbol{r}_l(N)\}$, each of which is a set of values over the index $l$, are arbitrarily chosen. The $N$ is chosen large enough that the steps are of short duration $\Delta t$. If I choose to add another step, Eq.(18) makes clear that another factor is required for each part of the formula, i.e.

(20)
$$\prod_{l=1}^{n} d^3\boldsymbol{r}_l(*)\left(\frac{1}{\sqrt{4\pi D_l \Delta t}}\right)^{3} exp[-D_l \Delta t \boldsymbol{\Lambda}_l \cdot \boldsymbol{\Lambda}_l]$$

The differential volume (the first factor) is always smaller than the reciprocal of the normalization factor (the second factor), and the exponential factor is less than one unless every Lagrange multiplier vanishes. Thus, adding a step to the Markov chain will lower the probability. The time interval, $\Delta t$, will also shorten if $t_{if}$ remains unchanged. If we add many more steps then the time interval, $\Delta t$, will definitely get smaller and the differential volume will have to be reduced to keep it less than the normalization factor. Only in this way will the tube of paths remain confined near the peak of the transition probability densities given in Eq.(5). We see that as the Markov chain approaches a continuous path the tube gets narrower and the probability goes to zero. This is no surprise since the measure for a continuous point-wise trajectory in coordinate space is zero.

The deterministic limit can be approached from a different perspective. Recursion formula Eq.(15) may be rewritten as

(21)
$$(r_l(j) - r_l(j-1) + \beta D_l \Delta t \nabla_l U(r^{3n}(N-1))) = -2D_l \Delta t \Lambda_l$$

or as

$$\frac{r_l(j) - r_l(j-1)}{\Delta t} = -\beta D_l \nabla_l U(r^{3n}(N-1)) - 2D_l \Lambda_l$$

As $\Delta t$ goes to zero, the spacing of the coordinate positions for adjacent steps also goes to zero and the limit is achieved.

(22)
$$\frac{d r_l}{dt} = -\beta D_l \nabla_l U - 2 D_l \Lambda_l$$

Each particle type is coupled to every other particle type, in principle, by the potential $U$. This deterministic equation represents the strongly overdamped dynamics that underlies the contacted Markov process in coordinate space. It is first order in the time derivative as befits a non-inertial heavily viscously damped motion. I emphasize this point because it helps to understand what the Lagrange multipliers will be doing.

Monotone relaxation to the local minima

It may happen that the initial point and the final point are separated by a barrier no matter what hypothetical path is taken. Or it may be the case that there are paths from the initial point to the final point that do not encounter a barrier. In the second case, the path will monotonically connect the initial and final points. The force term in Eq.(22) has a minus sign so that a positive force moves the point in the negative direction and *visa versa*. If the force vanishes then the motion abruptly stops, as is typical of overdamped motion but very unlike inertial motion. If the initial and final points are connected by a monotonic path, without barriers, then the dynamics will connect the points in a finite time. One exception exists if the final point happens to be a local minimum. The final approach to the local minimum will satisfy Eq.(22) with the force's coordinate dependence expanded around the local minimum. If the derivative of the force at the local minimum does not vanish the final approach is an exponential decay that takes infinitely long time to be reached. Even if the force derivative happens to vanish, and perhaps also higher derivatives too, the final approach stage will take infinite time (not exponentially in this case but algebraically). Thus local minima are critical points and we will have to deal with them in a special way. Note that just because an initial point can go to a final point without barriers this does not mean the reverse motion happens too. Deterministic motion without barriers that takes the initial point to the final point is expressed by Eq.(22) with all Lagrange vectors set equal to zero. To reverse the motion non-zero Lagrange vector values are required in order to overcome the unfavorable effect of the force term that is now acting like a barrier. If the force had been positive to create the original motion then the Lagrange vectors must be negative to create the reverse motion. Looking back at Eq.(19) we see that the probability for the reverse path will be much less than the probability for the forward path in this case because of the exponential factor. It is not impossible for the reverse motion to happen, it is just very low probability to happen. The molecular fluctuations in the system create the possibility of low probability events. These events are strictly forbidden by the deterministic dynamics unless sufficiently strong Lagrange multipliers are invoked. The multipliers popped up when we found the criterion for the most probable path and they represent the fact that the process is really stochastic. The fluctuations can be explicitly turned on in order to get to local minima in finite time or to get past an inflection point where the force and perhaps some of its derivatives happen to vanish.

## Fluctuations in the system

There are two ways to determine the fluctuations that match up with the deterministic dynamics. Eq.(1) can be viewed as the Fokker-Planck equation for coupled stochastic equations, so-called Langevin equations. These equations in the present context are

(23)
$$\frac{d\boldsymbol{r}_l}{dt} = -\beta D_l \nabla_l U + \widetilde{\boldsymbol{F}}_l(t)$$

wherein the $\widetilde{\boldsymbol{F}}_l(t)'s$ are Gaussian white noises with the two properties

(24)
$$<\widetilde{\boldsymbol{F}}_l(t)> = 0 \ \ and \ \ <\widetilde{\boldsymbol{F}}_{lj}(t)\widetilde{\boldsymbol{F}}_{l'k}(t')> = 2D_l \delta_{ll'} \delta(t-t') \delta_{jk}$$

wherein $j$ and $k$ are Cartesian component labels. The same result can be derived using functional calculus as we showed some time ago [17]. No Lagrange multipliers appear here. They arise when the most probable path is desired whereas these stochastic equations describe all possible paths. However, as already adumbrated above, the pathologies of overdamped motion at minima and inflection points that lead to infinite times can be avoided by explicitly using the fluctuations when the motion is within a standard deviation or two of these troublesome points.

## IV. The most probable paths

The characterization of the most probable path given above is for a Markov chain of $N$ steps each of which is for the short time interval $\Delta t$. All other paths of $N$ steps of time interval $\Delta t$ are less probable. Another perspective is gained by considering the entropy generated by the paths. From Eq.(7) we find that the path entropy, that is the logarithm of the path probability multiplied by the negative of Boltzmann's constant, is given by

(25)
$$S^F(path) = -k_B \sum_{j=1}^{N} \sum_{l=1}^{n} \left( \ln\left(\frac{d^3 \boldsymbol{r}_l(j)}{(4\pi D_l \Delta t)^{3/2}}\right) - \frac{(\boldsymbol{r}_l(j) - \boldsymbol{r}_l(j-1) + \beta D_l \Delta t \nabla_l U(\boldsymbol{r}^{3n}(j-1)))^2}{4 D_l \Delta t} \right)$$

The ratio in the logarithm terms is less than one and constant by construction and, therefore, generates a logarithmic value that is negative. Therefore, every term on the right-hand side of the equation is positive. In Eq.(8) we made sure that the sum of squares that also appears in Eq.(25) is a minimum for the most probable path. This means that any path other than the most probable path generates more entropy. This is the central result of the analysis: ***the most probable path generates the least entropy for a N step path of $\Delta t$ duration steps***. The global entropy can be written in terms of $R$ of Eqs.(1, 2) in the well known form $S = -k_B \int d^{3n} R \ln R$ but is not a state function at the level of point coordinates given in Eq.(25). Indeed, paths other than the most probable path that connect the same initial and final points generate more path entropy. On the other hand the internal energy is just $U$ and is a state function at the level of points in coordinate space. Thus for all paths connecting the initial point and the final point, whatever their probability, generate the same change in internal energy, $\Delta U = U(\{\boldsymbol{r}_l(N)\}) - U(\{\boldsymbol{r}_l(0)\})$. Therefore, we may also express the central result as we did in [7b]: *the most probable path dissipates the least Helmholtz free energy for a N step path of Δt duration steps*.

For application to the origin and evolution of life a subtle change in emphasis proves useful. The duration of each path considered in the minimization principle (least generation of entropy; least dissipation of Helmholtz free energy) is $t_{if} = N\Delta t$. Therefore the *rate* of entropy generation over the whole path is least for the most probable path. This means: *the most probable path has the least rate of entropy generation for a N step path of Δt duration steps*. This principle appears to depend on having chosen the context of coordinate space rather than that of full phase space [15, 16]. As was stated in [7b] using a contracted description in coordinate space conforms with what we see observationally, as opposed (*apposed*) to full phase space which includes momentum variables that are not observable because of their sub-picosecond changes caused by myriads of thermal, molecular collisions.

## V. Statistical thermodynamics of cellular life

The oxidation of glucose can take place rapidly and produce heat by simply burning in air (oxygen). This converts the available free energy into heat and does no useful work unless harnessed to a steam engine or some other converter of heat. Energy metabolism is the slow oxidation of glucose with many small steps each of which conserves some of the available free energy for other metabolic purposes including the synthesis of polymers. One might conclude that energy metabolism has slowed down the natural tendency for glucose to rapidly oxidize as a flame in order to harness some of the energy. However, glucose and oxygen do not spontaneously combust under normal conditions. An ignition catalyst is required, such as a match or electric spark. So we might instead view metabolism as making the initiation of the oxidation process possible. This seems even more plausible when you consider that cellular processes are immersed in water that is also a product of glucose oxidation.

Let us suppose that we start with a system of amino acids in aqueous solution mixed with pyrophosphate. As its name suggests pyrophosphate can be produced by heat. Pyrophosphate (P~P) can serve as a primitive analogue of ATP conceptually and chemically. Indeed, there are microorganisms that can use P~P in place of ATP in many different reactions [4] and even some, *Rhodospirrilum rubrum,* that putatively use it exclusively. With this simple mixture we can set up our basic problem. Laboratory experiments have shown that P~P and an amino acid can react to form *amino acyl carboxyl phosphate*, a primitive analogue of the modern activated amino acid, the amino acyl adenylate. When this happens, one phosphate of the original P~P ends up in an ester bond with the amino acid carboxyl group while the other phosphate ends up as monomeric orthophosphate. The amino acyl carboxyl phosphate is an activated species and two of them can react to make a dipeptide polymer (oligomer) of two amino acids. This polymer can subsequently react with another amino acyl phosphate to make a tripeptide and so on. As the peptide bond forms the carboxyl phosphate is released as orthophosphate and the carboxyl group forms an amide (peptide) bond with the amino group of the other amino acid. Eventually these peptides will hydrolyze and the final state will be a mixture of amino acid monomers and orthophosphate. This final state can be reached directly by hydrolysis of P~P without the reactions of amino acid activation, polymerization and final hydrolysis. Which path will happen more probably, the circuitous pathway of amino acid activation, polymerization and hydrolysis, or the direct pathway of hydrolysis of P~P without involvement of amino acids? One may surmise that the hydrolysis of P~P will take place rapidly and produce heat. This is in fact not so and the spontaneous hydrolysis of P~P in the absence of biological catalysts is extremely slow [18]. Specifically, at 25 ºC and pH 8.5 hydrolysis of $MgP$~$P^{2-}$ proceeds with a rate constant $2.8 \times 10^{-10} s^{-1}$, whereas *E. Coli* pyrophosphatase has a turnover number of 570 s$^{-1}$. This is a $2 \times 10^{12}$-fold enzymatic rate enhancement. Put another way, there is one natural, non-enzymatic, hydrolysis every 100 years! Although P~P hydrolysis is very favorable thermodynamically the natural rate is very small (it can vary over several orders of magnitude depending on pH and $Mg^{2+}$ concentration). Cellular life has had to evolve phosphatases, P~P hydrolysis catalyzers, in order to function at a reasonable rate. The interaction with amino acids described above may be an indirect way to enhance the natural slow rate of P~P hydrolysis. Wouldn't that make it less probable because the rate of path entropy would be greater?

It turns out that the rates for non-enzymatic hydrolysis of amino acyl carboxyl phosphates and of peptides are also slow. Simple dipeptides have been shown to have half-lives measured in hundreds of years [19]. Thus it is not simple to decide that one pathway generates entropy faster than another and is, therefore, less probable. A benefit of this slow rate of peptide hydrolysis is that plenty of time exists for feedback effects by the peptides on the reaction steps.

The context in which the principle for the most probable path was developed was one in which two points in coordinate space are selected and a fixed length of time was selected for the transition of the path trajectory from the initial point to the final point. Of all paths between the two points in the specified time the most probable generated the least path entropy. The path entropy is not the macroscopic system entropy that depends on concentrations variables and not on point states. Nor is it the activation entropy of *transition state theory*. The latter is not defined for coordinate space points because it is not a state function on points, unlike the internal energy that is well defined on points by $U$ of Eq.(3). The macroscopic entropy depends on concentration variables that represent regions of coordinate space that are made up of points that are equivalent in the sense that all of these points correspond to a given number of each type of molecule, randomly distributed in space. Our initial state of P~P molecules and amino acids makes up a complicated sub-region of all of coordinate space, as does the final state of orthophosphate and amino acids. We may take any point from the initial region and follow it to any point in the final region. For all of these possible pairs of points the paths may correspond to direct hydrolysis of P~P or indirect hydrolysis coupled to amino acid polymerization and hydrolysis.

In reality, quantum mechanics is required to correctly describe chemical bonds and reactions. In the present context we model these events through classical potentials having barriers protecting deep attractive wells that mimic bonding, possibly requiring three body potentials or even higher order many body potentials. Thus our system will have a fixed number of atoms of various types that can be found associated with each other through deep potential well attractions as though they were molecules. This means that all of the coordinates are assigned at the outset and are not created or destroyed as molecular combinations come and go. Otherwise the adaptive computational problem becomes very complex as new molecules arise and require their own new coordinates [20]. One additional complication is that the potentials, so far imagined to be dependent on the radial distance between atoms only, may depend on the angles of orientation as well. Another is that water molecules are involved in both dehydrations and hydrolyses but we have removed water molecules from explicit consideration in favor of Brownian motion on all other species. By including enough hydrogen and oxygen atoms in the initial mixture water molecules can form and disappear along with other molecules such as amino acids and phosphates through the interactions implicit in the deep wells of the classical potentials.

I will now attempt to cut through all these complications and get to the heart of the matter. Either a large amount of free energy is released directly in one step, or several small steps result in the same end state instead. Observation tells us that the one step process is intrinsically very slow. These steps involve highly correlated groups of atoms, "molecules," and the key reactive atoms, oxygen, nitrogen and phosphorus, are bearers of the coordinates seen in Eq.(22). Since in hydrolysis the *nucleophilic attack* of a P~P P by a water O is very slow the barrier is high and the $\Lambda_l$ is large for some particular $l$. In Eq.(20) this means there are factors of $exp[-D_l \Delta t \Lambda_l \cdot \Lambda_l]$ for this $l$. Now suppose the same barrier is overcome by three smaller steps, each about a third as high a barrier. These barriers are of different heights but may be of comparable widths as is often tacitly assumed in transition state theory diagrams. Thus the slope of the high barrier is roughly three times the slope of each of the smaller barriers. The quantity $\Lambda_l$ corresponds with this slope. In our model situation a carboxyl O attacks a P~P P and a amino acid N attacks a carboxyl phosphate P and finally a water O attacks a peptide amide bond. This introduces the factors $exp[-D_a \Delta t \Lambda_a \cdot \Lambda_a]$, $exp[-D_b \Delta t \Lambda_b \cdot \Lambda_b]$ and $exp[-D_c \Delta t \Lambda_c \cdot \Lambda_c]$. Each $\Lambda_{a,b,c}$ is perhaps 1/3 $\Lambda_l$. Therefore we compare $exp[-D_l \Delta t \Lambda_l \cdot \Lambda_l]$ with

(26)
$$exp[-D_a \Delta t \Lambda_a \cdot \Lambda_a] exp[-D_b \Delta t \Lambda_b \cdot \Lambda_b] exp[-D_c \Delta t \Lambda_c \cdot \Lambda_c] =$$
$$exp\left[-3 D_l \Delta t \frac{1}{9} \Lambda_l \cdot \Lambda_l\right] = exp\left[-\frac{1}{3} D_l \Delta t \Lambda_l \cdot \Lambda_l\right]$$

If this reasoning is correct then the three steps process generates less path entropy than is generated by the one step process. It is more probable for P~P to generate peptides then to directly hydrolyze. The emergent feedback properties of the transient peptides usher in many evolutionary possibilities. (Note that if the three lesser slopes are not each 1/3 of the larger slope but instead are given by three numbers, a, b, and c, each inside the open interval (0, 1), and added together equal to 1, then the coefficient in the last equality of Eq.(26), that is 1/3, becomes

(27)
$$(a^2 + b^2 + c^2) < 1$$

This is sufficient for the conclusion above to still hold.)

A similar analysis of what is the fate of nucleotide triphosphates, direct hydrolysis or polynucleotides, is possible. This question focuses on the phosphodiester bonds of a nucleotide triphosphate or on the ribophosphate bond of a polynucleotide. Polymerization of nucleotide triphosphates releases P~P and forms ribophosphate bonds. These eventually hydrolyze yielding nucleosides and orthophosphate. The P~P also eventually hydrolyses. On the other hand nucleotide triphosphates can hydrolyze yielding nucleosides and orthophosphate directly. Which process is more probable is not easy to decide in the absence of good data on the uncatalyzed rates. The formation of the ribophosphate bond during polymerization preserves some of the free energy of the triphosphate that is cleaved during this reaction. This parallels the situation in peptide formation. Perhaps the barrier analysis is similar. If polymerization is not more probable than hydrolysis then catalytic peptide interactions favoring polymerization may be necessary for RNA to have arisen. In the contemporary mechanism of polynucleotide polymerization many proteins are involved. For that matter, many polynucleotides are involved in protein synthesis.

While the arguments given above may have validity for the beginnings of life there remain problems later on in evolution that have to do with chemical particulars. The most fundamental one is how the *amino acyl tRNA synthetases* evolved and locked in the genetic code. I have explored this question elsewhere [21]. In that exposition I also present my view on how an RNA world emerged and generated protein biosynthesis, the problem posed in the sub-section on energy metabolism and the monomer to polymer transition in section **I**.

Ronald F. Fox
December 1, 2015
Smyrna, Georgia